\documentclass[seceq,preprint]{ptptex}

\usepackage{graphicx}
\usepackage{eucal}

\preprintnumber[3cm]{
RIKEN-TH-207\\April, 2012}

\title{
Matrix model criticality and resonant tunneling%
}

\author{
Tsukasa \textsc{Tada}%
}

\inst{
RIKEN Nishina Center for Accelerator-based Science\\
Hirosawa 2-1, Wako, Saitama 351-0198, Japan
}

\abst{
We suggest that the Hermitian matrix models with resonant tunneling  may exhibit novel criticality. Some features of the proposed criticality are explored. In particular, we argue that the new critical point is connected with the first-order transition.
}

\begin{document}

\maketitle

\section{Introduction}
The tunneling phenomenon is a salient feature of quantum physics that even physicists sometimes find peculiar. One such example of it is resonant tunneling. The tunneling effect is usually characterized by an exponential suppression by a potential barrier; however, if a pair of potential barriers exists and certain conditions are met, tunneling occurs with a probability of unity, as we will explain in the next section.

Resonant tunneling has been extensively studied, but interesting applications and researches of it still exist. This paper presents one such application\cite{Tye:2011xp}\cite{Tye:2009rb}\cite{Saffin:2008vi}\cite{Copeland:2007qf}\cite{Sarangi:2007jb}. In the following sections, we investigate resonant tunneling in the context of matrix models and explore its consequences. 

Matrix models may seem an unlikely arena for the tunneling phenomenon because, first, quantum fluctuations in the models described by $N \times N$ matrices are suppressed  when one takes a large-$N$ limit, in which almost all the useful analyses can be conducted. There, the system usually stays at a stable vacuum whose nature is well understood, until a change in the parameters sets some of the degrees of freedom unstable and causes a divergence. This is the critical point of usual matrix models.  Our point is that new instabilities caused by resonant tunneling unfold, while ordinary tunneling effects are much subtle in matrix models \footnote{The tunneling  of an eigenvalue of matrix models is attributed to non-perturbative effect of non-critical string theory \cite{Shenker:1990uf,Eynard:1992sg,David:1990sk,David:1992za,Lechtenfeld:1991kc}. It can be understood as a non-critical version of D-branes\cite{Polchinski:1994fq, Zamolodchikov:2001ah,Fukuma:1999tj,Neves:1997xt}\cite{Klebanov:2003km,Martinec:2003ka,McGreevy:2003kb,Alexandrov:2003nn}. This perspective prompted further investigations \cite{Takayanagi:2003sm,Douglas:2003up,Hanada:2004im}.}.

Since we need to incorporate the tunneling phenomenon, our current analysis is, among other various matrix models, limited to the one-dimensional one, where tunneling effect is most straightforward to calculate. The other cases are left for future study.

\section{Resonant tunneling}\label{rt}
\begin{figure}[htbp]
\begin{center}
\includegraphics[width=12cm]{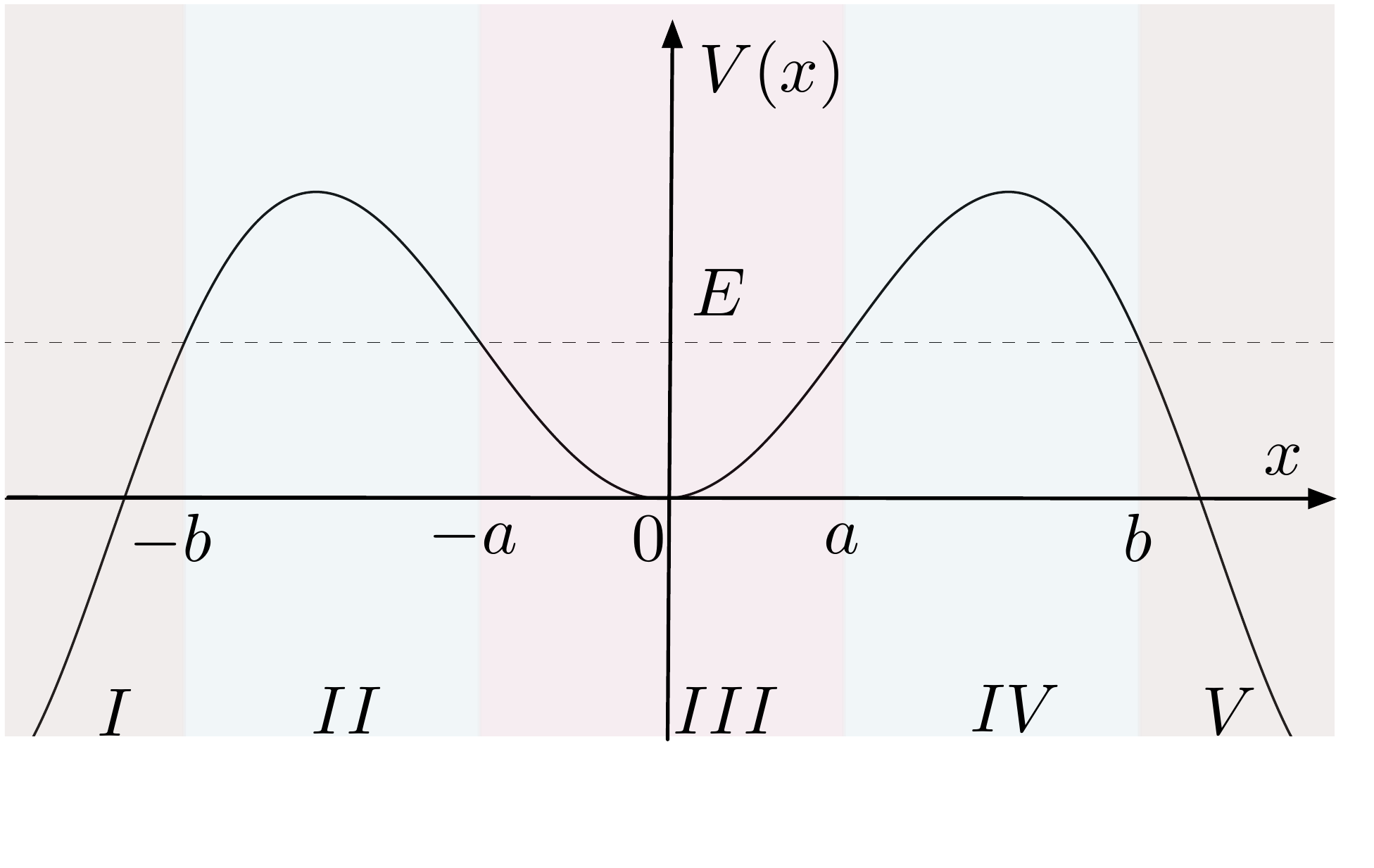}
\caption{Potential $V(x)$ divided into  five regions}
\label{fig}
\end{center}
\end{figure}

Let us first illustrate resonant tunneling\cite{Merzbacher:1997} \cite{Bohm:1989} . Consider the one-dimensional Schr\"odinger equation
\begin{equation}
\frac{d^2 \psi}{dx^2}+\frac{2m(E-V(x))}{\hbar^2}\psi=0,
\end{equation}
where, for the sake of simplicity,  we take $V(x)$ to be symmetric under the reflection at the origin $x \leftrightarrow -x$ as shown in  Fig. \ref{fig}.
Now suppose a situation in which an incident wave enters from the left at $x<-b$ (region $I$ in Fig. \ref{fig}) and eventually exits from the  right at $x>b$ (region $V$ in Fig. \ref{fig}.). Then, within the WKB (Wentzel-Kramers-Brillouin) approximation, the incident wave function in region $I$ can be expressed as
\begin{equation}
\psi_{I}= \frac{A}{k} e^{i \int^x_{-b}k dx} +\frac{B}{k} e^{-i \int^x_{-b}k dx}, 
\end{equation}
where
\begin{equation}
 k(x)\equiv \sqrt{\frac{2m}{\hbar}(E-V(x))}.
\end{equation}
The outgoing wave function in the region $V$ is given by
\begin{equation}
\psi_{V}=\frac{F}{k} e^{i \int^x_{a}k dx} ,
\end{equation}
and the wave function in region $III$ $(-a < x <a)$ is given by
\begin{equation}
\psi_{III}= \frac{C}{k} e^{i \int^x_{a}k dx} +\frac{D}{k} e^{-i \int^x_{a}k dx}
.
\end{equation}
The relationships among the coefficients $A, B, \dots , F$ in the above expressions  can be calculated using the connection formula. In fact, these relationships can be conveniently cast into a matrix form; for example,
\begin{equation}
\left(\begin{array}{c}C \\D\end{array}\right) =\frac12 \left(\begin{array}{cc}2\theta+\frac{1}{2\theta} & i(2\theta-\frac{1}{2\theta}) \\-i(2\theta-\frac{1}{2\theta}) & 2\theta+\frac{1}{2\theta}\end{array}\right)\left(\begin{array}{c}F \\0\end{array}\right),
\end{equation}
where
\begin{equation}
\theta=e^{\int^b_a \kappa dx}, \ \ \kappa(x)\equiv\sqrt{\frac{2m}{\hbar}\left(V\left(x\right)-E\right)}. \label{thetadef}
\end{equation}
Then, the  transmission coefficient between regions $III$ and $V$ can be determined as follows:
\begin{equation}
T_{III\rightarrow V}=\frac{|F|^2}{|C|^2}= \frac{4}{\left(2\theta +\frac{1}{2\theta}\right)^2}.
\end{equation}
Note that $\theta$ contains the exponential factor, as shown in Eq. (\ref{thetadef}); therefore, the transmission from regions $III$ to $V$ is exponentially suppressed as expected. In addition, one can apply the connection formula between regions $I$ and $III$ and combine the result with the above calculation to obtain
\begin{eqnarray}
\left(\begin{array}{c}A \\B\end{array}\right)&=&\frac12 \left(\begin{array}{cc}2\theta+\frac{1}{2\theta} & i(2\theta-\frac{1}{2\theta}) \\-i(2\theta-\frac{1}{2\theta}) & 2\theta+\frac{1}{2\theta}\end{array}\right)
\left(\begin{array}{c}e^{-\frac{J}{2\hbar}i}\frac12\left(2\theta+\frac{1}{2\theta}\right) F \\-ie^{\frac{J}{2\hbar}}\frac12\left(2\theta-\frac{1}{2\theta}\right)F\end{array}\right),
\\
&=&\frac{F}{4}\left(\begin{array}{c}2\cos \frac{J}{2\hbar} \left(4\theta^2+\frac{1}{4\theta^2}\right)-4i\sin \frac{J}{2\hbar} \\-2i\cos \frac{J}{2\hbar} \left( 4\theta^2-\frac{1}{4\theta^2}\right)\end{array}\right),
\end{eqnarray}
where
\begin{equation}
J\equiv 2\int^a_{-a}\sqrt{\frac{2m}{\hbar}(E-V(x))} dx =\oint p dq.\label{Jdef}
\end{equation}
Then, the total transmission coefficient is
\begin{equation}
T_{I\rightarrow V}=\frac{|F|^2}{|A|^2}= \frac{4}{\left(4\theta^2 +\frac{1}{4\theta^2}\right)^2\cos^2 \frac{J}{2\hbar}+4\sin^2\frac{J}{2\hbar}}. \label{totaltransmission}
\end{equation}

Note that the transmission coefficient given in Eq. (\ref{totaltransmission}) also exhibits an exponential suppression through its $\theta$ dependence. However, a special situation occurs if the following condition is met:
\begin{equation}
J=2\pi\hbar\left(n+\frac12\right), \ \ n \hbox{: integer}. \label{Jcond}
\end{equation}
In this case, the cosine in the denominator of Eq. (\ref{totaltransmission}) vanishes and the sine becomes unity, yielding a transmission coefficient $T_{I\rightarrow V}$ of exactly unity. The exponential suppression of tunneling vanishes completely, and the transmission occurs with a probability of unity. This remarkable phenomenon is known as resonant tunneling \cite{Merzbacher:1997}\cite{Bohm:1989}.

\section{Matrix model}
\begin{figure}[htbp]
\begin{center}
\includegraphics[width=12cm]{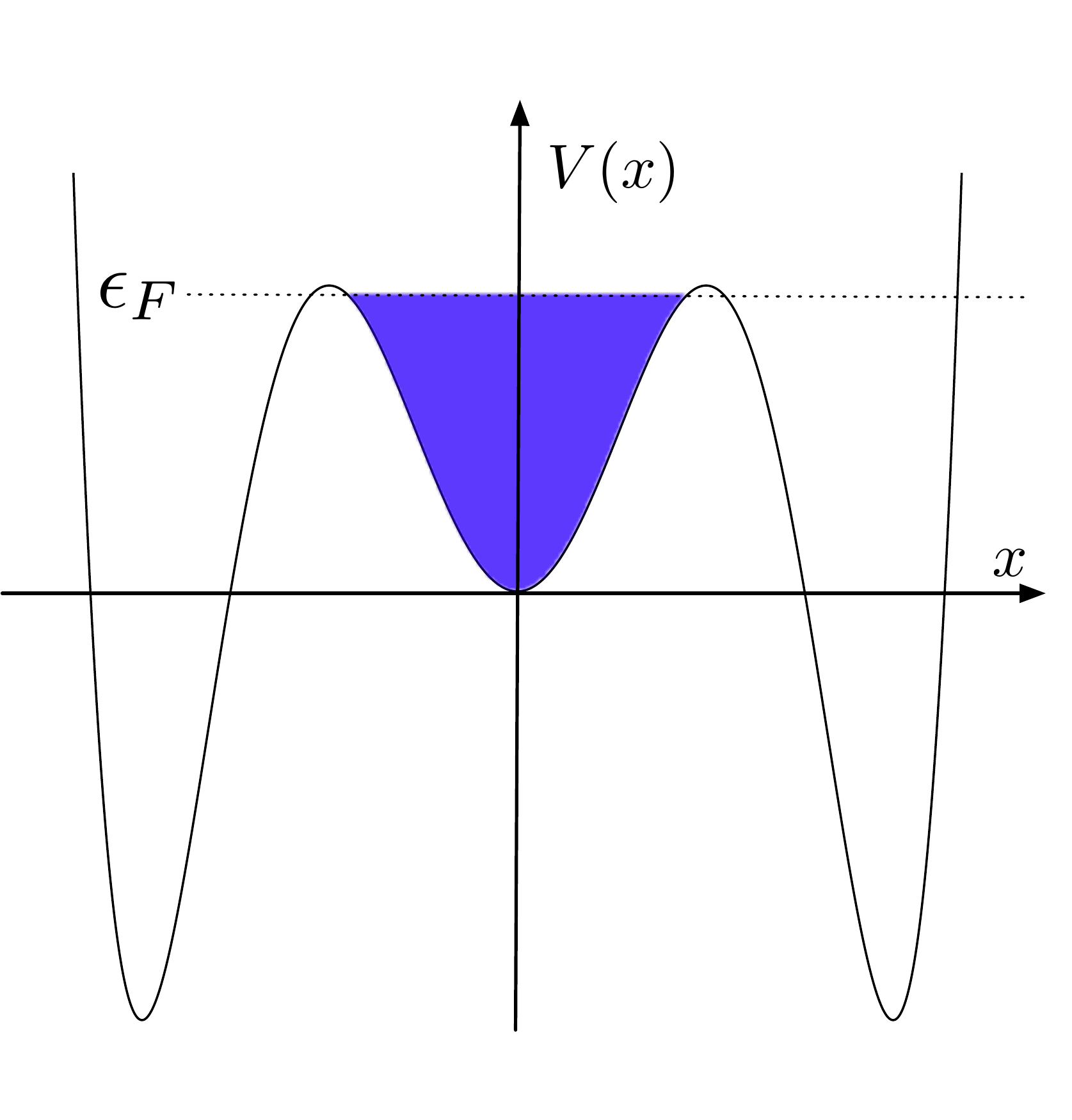}
\caption{Fermions filling the energy levels.}
\label{fig2}
\end{center}
\end{figure}

In the previous section, we treated a single-particle wave function. Now we consider the case for $N^2$ particles, each of mass $m$. Various possible interactions among these $N^2$ particles may exist, but we choose the  interactions and potential such that all the degrees of freedom can be cast into an $N\times N$ Hermite matrix $M_{ij}$  and the Hamiltonian $H$ takes the following form:
\begin{eqnarray}
H&=&-\frac{\hbar^2}{2m} \Delta_M + \hbox{Tr} V(M) , \nonumber \\
&\Delta_M  & =\sum_{1 \leqslant i \leqslant N} \frac {\partial^2}{\partial M_{ii}^2} +\frac12 \sum_{1 \leqslant i <j \leqslant N} \Biggl[ \frac {\partial^2}{\partial Re M_{ij}^2}+\frac {\partial^2}{\partial Im M_{ij}^2}\Biggr]  \label{mm} \\
&V(M)&=\frac12  M^2 + \sum_p g_p N^{1-\frac{p}{2}} M^p.\nonumber
\end{eqnarray}
For the sake of simplicity, $\hbar$ is set to unity here. In their seminal paper \cite{Brezin:1977sv},  Brezin, Itzykson, Parisi and Zuber made an ingenious observation that because of invariance under the $U(N)$ transformation $M \rightarrow UMU^{-1}$, the ground state of the above $N^2$ particles, which is assumed to possess  the invariance just mentioned, is governed by the following Schr\"odinger equation for N eigenvalues of $M$:
\begin{equation}
\sum_{1 \leqslant i \leqslant N} \Bigl\{ -\frac{\hbar^2}{2m} \frac {\partial^2}{\partial \lambda_{i}^2} + \frac12\lambda_i^2 + \sum_p g_p N^{1-\frac{p}{2}}\lambda_i^p \Bigr\} \phi_N(\{\lambda_i\})=N^2E_{g}\phi_N(\{\lambda_i\}). \label{Schreq}
\end{equation}
Here, the scale of $E_{g}$ (the total energy of the system) is set in accordance with the total number of the degrees of freedom $N^2$, and the subscript indicates its dependence on the potential variables $\{g_p\}$. Considering the Van der Monde determinant, the above wave function $\phi(\lambda)$ must be completely antisymmetric among its $N$ variables; hence, Eq. (\ref{Schreq}) represents $N$ free fermions under the following potential:
\begin{equation}
V(\lambda)=\frac12\lambda^2+\sum_p g_p N^{1-\frac{p}{2}}\lambda^p.
\end{equation}
Thus, the physics of the sector of interest is governed by N free fermions, each of which is subject to the following Schr\"odinger equation:
\begin{equation}
\frac{d^2\phi}{dx^2}+\frac{2m}{\hbar^2}\left( N\epsilon -V\left( x\right) \right)\phi=0,
\end{equation}
where the scale of $\epsilon$ (the energy of each fermion) is again set in accordance with the number of the degrees of freedom, because each fermion carries approximately $N$ times the energy of the original particle.
Then, the ground state of the whole system is easy to construct by filling the energy levels from the bottom up to the Fermi energy $\epsilon_F$ (Fig. \ref{fig2}).
$\epsilon_F$ is determined from knowing that the total number of fermions is $N$. For the large $N$ approximation, the following equation holds:
\begin{equation}
N=\int \frac{d\lambda dp}{2\pi \hbar}\Theta \left( \epsilon_F-\frac{p^2}{2m} -V\left(\lambda\right) \right), \label{NeF}
\end{equation}
where $\Theta$ denotes the step function. The integral in Eq. (\ref{NeF}) implicitly determines $\epsilon_F$ but a singularity exists when the level of $\epsilon_F$ reaches any local maximum of the potential $V(\lambda)$. This singularity corresponds to the critical point where the present matrix model given in Eq. (\ref{mm}) becomes equivalent to the $D=2$ non-critical string theory in the so-called double-scaling limit \cite{c1doublescaling}. The physics behind this criticality is that beyond that critical point, a fermion with the Fermi energy is no longer confined in the valley formed by the local maxima, and it can cross over one of them. Therefore, the system described by these fermions goes through a phase transition. Details of this critical behavior have been investigated \cite{Kazakov:1988ch}.

\section{New criticality through resonant tunneling}
\begin{figure}[htbp]
\begin{center}
\includegraphics[width=12cm]{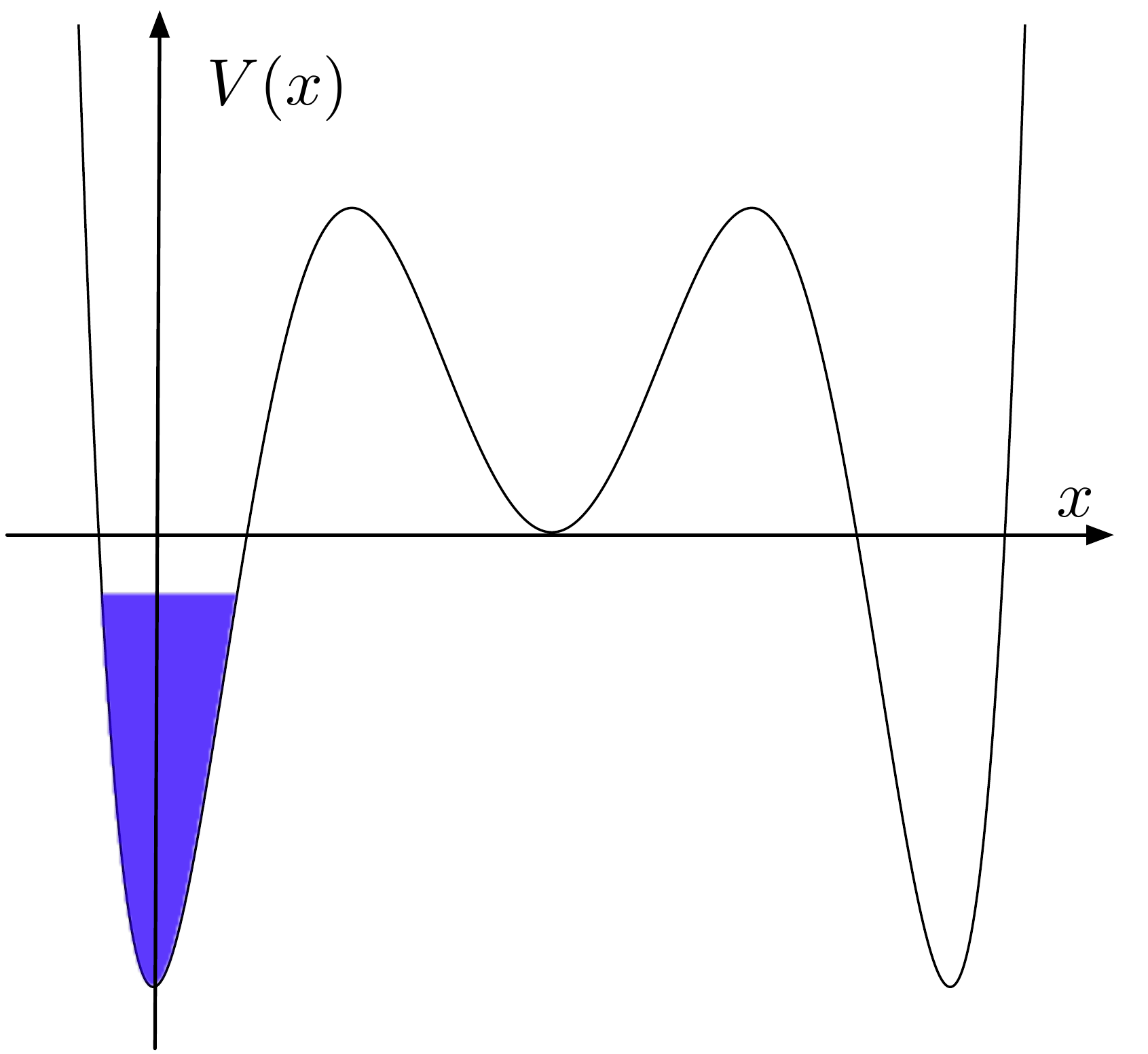}
\caption{Energy levels starting at the left minimum are filled up to the Fermi energy.}
\label{fig3}
\end{center}
\end{figure}

Now we argue that the matrix model presented in Eq. (\ref{mm}) exhibits a  different kind of criticality than that explained in the previous section if resonant tunneling is considered. Consider a case with the potential shown in Fig. \ref{fig3}.
The energy levels that corresponds to the motion near the origin are the first to be filled, and they are filled up to the Fermi energy $\epsilon_{F}$, which is determined using Eq. (\ref{NeF}). Because different values of the parameter $g_{p}$ in the potential lead to different values of the Fermi energy $\epsilon_{F}$, one can tune the level of $\epsilon_{F}$ by gradually varying the value $g_{p}$. 

\begin{figure}[htbp]
\begin{center}
\includegraphics[width=12cm]{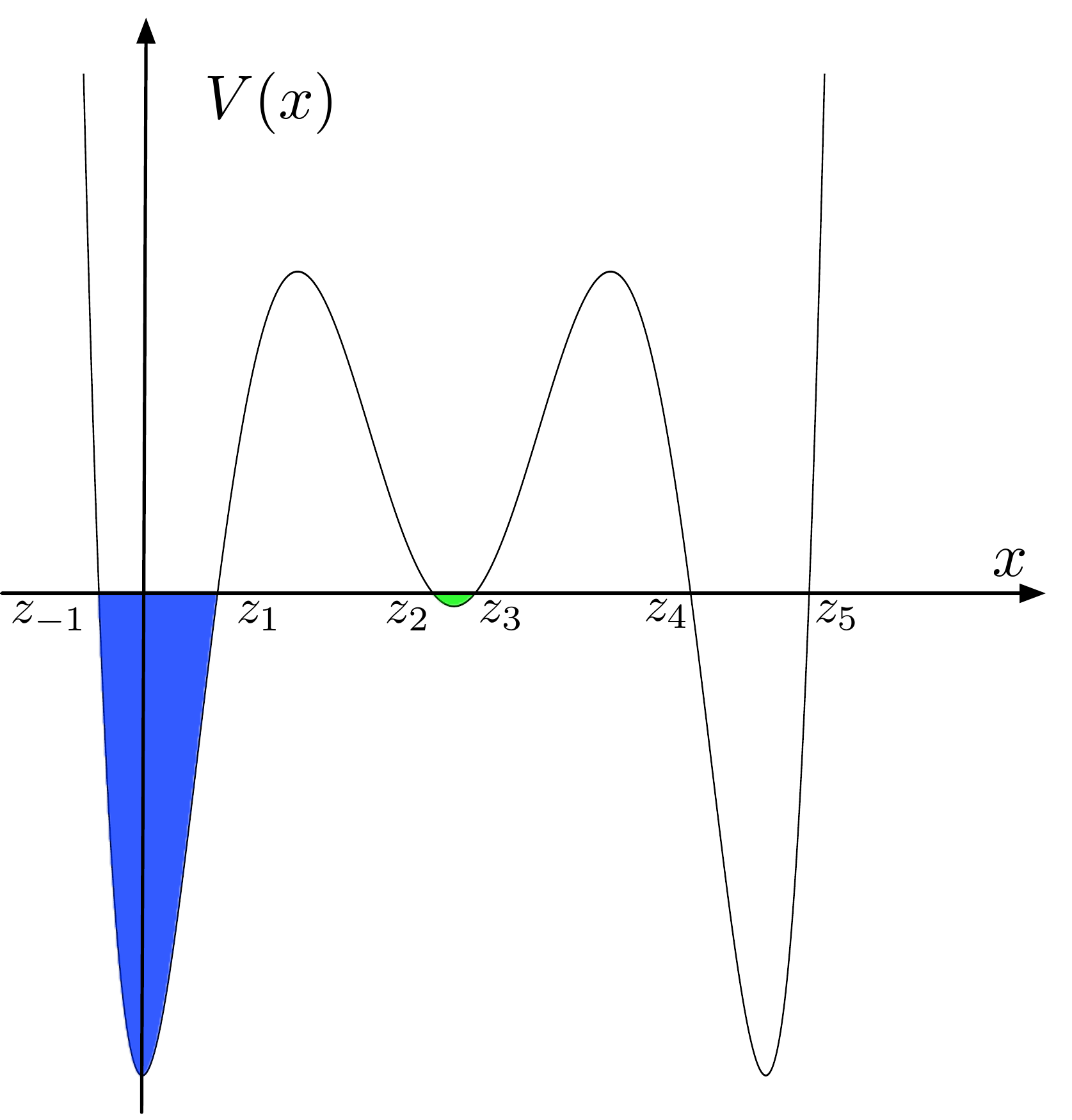}
\caption{Resonant Tunneling. The region shaded in green is relevant for $J$.}
\label{fig4}
\end{center}
\end{figure}

\begin{figure}[htbp]
\begin{center}
\includegraphics[width=12cm]{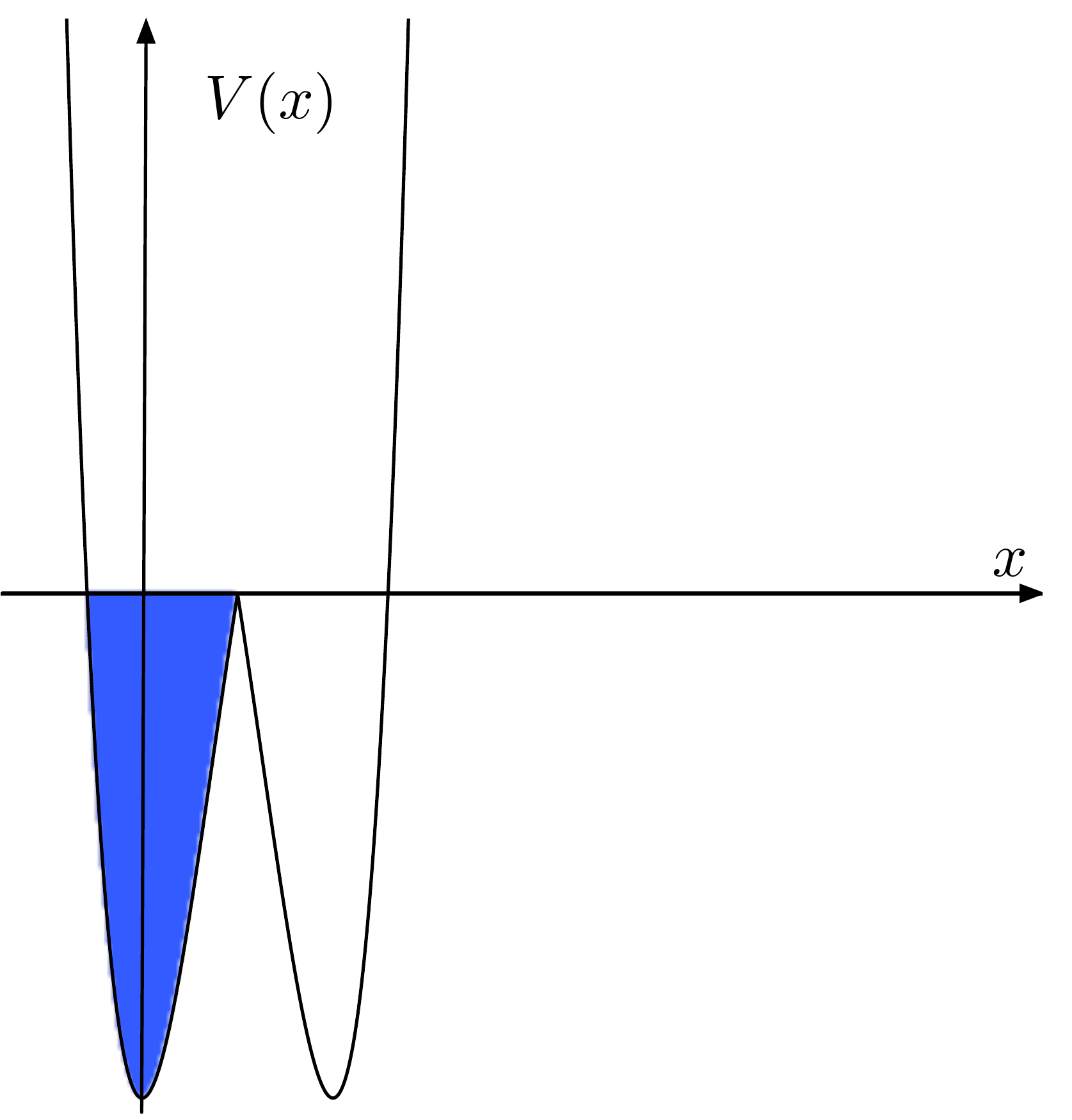}
\caption{Potential felt by fermions due to resonant tunneling}
\label{fig45}
\end{center}
\end{figure}

Figure \ref{fig4} depicts a case in which $\epsilon_{F}$ is zero and the central local minimum is slightly negative. Note that the shape of the potential is completely symmetric under reflection at the central local minimum, and the Fermi energy $\epsilon_{F}$ is less than the local maxima.
One can calculate a quantity similar to that in Eq. (\ref{Jdef}) in Section \ref{rt} where we discussed resonant tunneling:
\begin{equation}
J\equiv 2\int^{z_{2}}_{z_{3}}\sqrt{\frac{2m}{\hbar}(-V(x))} dx,
\end{equation}
where $z_{2}$ and $z_{3}$
are the nearest zero points of the potential to the local minimum. Now suppose $J$ satisfies the following:
\begin{equation}
J=\pi\hbar,\label{J0}
\end{equation}
which is nothing but the condition for resonant tunneling given in Eq. (\ref{Jcond}) with $n=0$. Therefore, the analysis described in Section \ref{rt} indicates that a fermion with energy $E=0$ can penetrate and pass through  two of the potential barriers into the other potential minimum region with a transmission coefficient of unity. From the viewpoint of the fermions, the Fermi energy gradually increases to zero as the potential parameters change, and if the potential was set to simultaneously satisfy Eq. (\ref{J0}), the potential barrier formed by the twin peaks suddenly disappears (Fig. \ref{fig45}).
Therefore,  the situation is similar to the criticality discussed in the previous section in the sense that the fermion with the highest energy crosses the potential barrier and senses the potential beyond the barrier. We claim that this is a novel criticality realized by the resonant tunneling phenomenon.

\begin{figure}[htbp]
\begin{center}
\includegraphics[width=12cm]{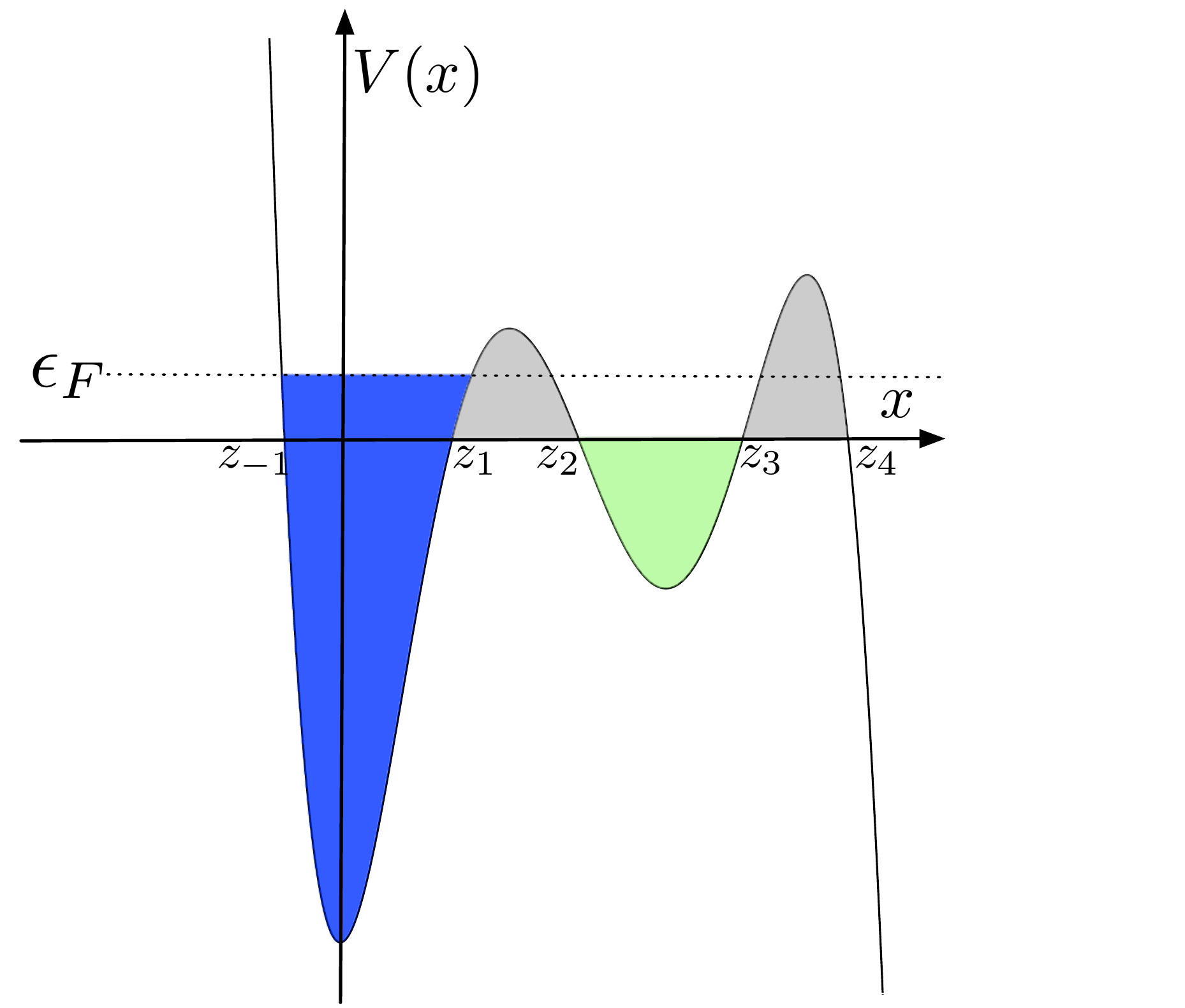}
\caption{More general potential}
\label{fig5}
\end{center}
\end{figure}

The novelty here is not only the association with the resonant tunneling. Let us introduce a more general potential such as the one shown in Figure \ref{fig5}.
In this case, the Fermi energy can assume a positive value without hitting the criticality, even if the potential satisfies the condition for the resonant tunneling:
\begin{equation}
J=2\int^{z_{3}}_{z_{2}}\sqrt{\frac{2m}{\hbar}(-V(x))} dx =2\pi\hbar\left(n+\frac12\right), \ \ n \hbox{: integer}. \label{Jcondnew}
\end{equation}
This is because for a general potential, the transmission coefficient takes the following form instead of that presented in Eq. (\ref{totaltransmission}):
\begin{equation}
T= \frac{4}{\left(4\theta \theta' +\frac{1}{4\theta\theta'}\right)^2\cos^2 \frac{J}{2\hbar}+\left( \frac{\theta}{\theta'}+\frac{\theta'}{\theta} \right)^2\sin^2\frac{J}{2\hbar}},
\end{equation}
where
\begin{equation}
\theta=e^{\int^{z_1}_{z_2} \kappa dx}, \ \ \kappa(x)\equiv\sqrt{\frac{2m}{\hbar}V\left(x\right)}, 
\end{equation}
and
\begin{equation}
\theta'=e^{\int^{z_3}_{z_4} \kappa dx}, \ \ \kappa(x)\equiv\sqrt{\frac{2m}{\hbar}V\left(x\right)},
\end{equation}
respectively. When the condition for resonant tunneling presented in Eq. (\ref{Jcondnew}) is satisfied, the transmission coefficient is
\begin{equation}
T= \frac{4}{\left( \frac{\theta}{\theta'}+\frac{\theta'}{\theta} \right)^2}.
\end{equation}
The transmission coefficient $T$ would be unity only if $\theta=\theta'$, otherwise it would exhibit an exponential suppression arising from $\theta$ and $\theta'$.
Now suppose that the potential in Fig. \ref{fig5} is tuned so that $\theta=\theta'$. Then, fermions with  energy $0<E<\epsilon_F$ could freely travel to the region $x>z_4$. Therefore, it would appear that the energy carried by such fermions escapes from the system instantaneously. One may interpret the escaped energy as latent heat. This observation suggests that the criticality we  introduced corresponds to the first-order transition rather than a second-order one. \footnote{We owe this observation to H. Kawai.}

\section{Conclusion}
In this paper, we demonstrated that the Hermitian matrix models with resonant tunneling could exhibit a new criticality. We suggested that this new criticality corresponds to the first-order transition, in contrast to the conventional critical point that corresponds to a second-order transition. While it is well established that the criticality associated with the second-order transition manifests $c=1$ non-critical string, the physical implications of the present criticality remains to be explored. In particular,  future studies could examine the detailed behavior of this new criticality.

\section*{Acknowledgements}
The author is indebted to H. Kawai for illuminating discussions and many insights provided towards the present paper. The author would like to thank S. H. Tye for bringing the subject  of resonant tunneling to his attention and giving a clear explanation of it. This work is supported in part by the Basic Science Interdisciplinary Project "Study on the Genesis of Matter."

%

\end{document}